\documentstyle[12pt]{article}
\begin{document}

\centerline{\bf COMPARATIVE STUDY OF THE ADIABATIC EVOLUTION OF }

\centerline {\bf A NONLINEAR DAMPED OSCILLATOR AND AN HAMILTONIAN}

\centerline {\bf  GENERALIZED NONLINEAR OSCILLATOR}

\bigskip
\bigskip
\centerline { O.V. Usatenko$^{\dagger *}$, J.-P. Provost$^{\dagger}$, G.
Vall\'ee$^{\ddagger}$ and
A. Boudine$^{\dagger \dagger} $} \bigskip
\bigskip
\bigskip
\bigskip
\centerline { PACS numbers $03.20.+i$}
\centerline {AMS Classification Scheme numbers 70 H, 34 C}
\bigskip
\bigskip
\bigskip
\bigskip
\bigskip
\bigskip
\bigskip

\noindent * On leave from the Department of Physics, Kharkov State
University, Kharkov 310077, Ukraine.
\medskip
\noindent $\dagger \*$ Institut Non Lin\'eaire de Nice, UMR CNRS 129,
Universit\'e de Nice Sophia
Antipolis, 1361 Route des Lucioles 06560 Valbonne, France.
\medskip
\noindent $\ddagger$  Laboratoire de
Math\'ematiques J.A. Dieudonn\'e, UMR CNRS 6621, Universit\'e de Nice
Sophia Antipolis,
Parc Valrose, 06108 Nice Cedex 2, France.
\medskip
\noindent $\dagger \dagger $ Departenent de Physique Theorique. Universite
de Constantine. 25000
Constantine. Algerie.

\bigskip

\eject
\vglue 7 cm

\noindent Abstract

In this paper we study to what extent the
canonical equivalence and the identity of the geometric phases of
dissipative and conservative linear
oscillators, established in a preceeding paper, can be generalized to
nonlinear ones. Considering
first the 1-D quartic generalized oscillator we determine, by means of a
perturbative time dependent
technic of reduction to normal forms, the canonical transformations which
lead to the adiabatic
invariant of the system and to the first order non linear correction to its
Hannay angle. Then,
applying the same transformations to the 1-D quartic damped oscillator we
show that this oscillator
is canonically equivalent to the linear generalized harmonic oscillator for
finite values of the
damping parameter (which implies no correction to the linear Hannay angle)
whereas, in an appropriate
weak damping limit, it becomes equivalent to the quartic generalized
oscillator (which implies a
non linear correction to this angle) .

\vfill \eject

 \noindent  INTRODUCTION
\medskip
The Hannay's angle [1] (classical counterpart of the Berry's geometric
phase [2]), originally
associated with the adiabatic evolution of classical hamiltonian systems,
has been recently extended to a large
class of dynamical equations corresponding to dissipative systems: non
linear equations with
limit cycles [3] or with more general internal symmetries [4], equations
describing the dynamics of the laser
[5], etc...In this context we have shown in a recent paper [6] that the
simplest dissipative system, namely the damped
harmonic oscillator specified by the time dependent Lagrangian
$${\cal L}(q, \dot q, \vec \mu) = {1\over 2}\, e^{2\!
\int ^t\!  \lambda (s) \, ds }\, ({\dot q}^2 - \omega _o^2 q^2 )  \eqno (1)$$
is canonically equivalent, even for time dependent parameters $(\lambda
,\omega_o) \equiv \vec \mu$,
to the generalized harmonic oscillator, a conservative system specified by
the Hamiltonian
$$H(P,Q,\vec \mu) = {P^2 \over {2}} + \lambda P Q + {{ \omega _o^2} \over
2} Q^2 \  .\eqno (2)$$
(The generating function of the canonical transformation is
 $F(q,P,t) = q P\,  e^{\int ^t   \lambda (s) \, ds }$).
As a consequence the Hannay's angles of the two systems are identical.
Their expression can be simply
recovered, like in [7], using the transformed variable
  $$ Q = {{\partial F(q,P,t)} \over {\partial P}} = q \,  e^{\int ^t
\lambda (s) \, ds } \eqno (3)$$
which brings the Lagrangian (1) to the form
 $$L(Q,\dot Q, \vec \mu) = {1 \over 2}({\dot Q}^2 - 2 \lambda Q \dot Q
-\omega^2 Q^2 ) \ \ \ \ \ \ \ \ \  (\omega^2 = \omega_o^2 - \lambda^2)
\eqno (4)$$
 which also reads
 $$L(Q,\dot Q, \vec \mu) = {1\over 2}({\dot Q}^2
-(\omega^2 - \dot \lambda ) Q^2 ) -{1\over 2} {d\over dt} (\lambda Q^2)\  .
\eqno (5) $$
In this later expression the
quantity $(\omega^2 - \dot \lambda )$ is the square of the
instantaneous frequency of the system. In the adiabatic limit where the
parameters
$\vec \mu$ are  slowly time-varying functions ${\vec \mu} (\epsilon t)$
(with $\displaystyle
{{\epsilon \over \omega_o} \ll 1}$) one can make a first order expansion of this
instantaneous frequency with respect to the small adiabatic parameter
$\epsilon$. One gets the well
known result [1,2]
 $$ \dot \Theta
= \omega - {{\dot \lambda} \over {2\omega}}\  . \eqno (6) $$
where the `dynamical' part $\omega$ of the time derivative $\dot \Theta$ of
the phase of the
oscillator appears to be corrected by an adiabatic contribution
$\displaystyle {-{{\dot \lambda}
\over {2\omega}}}$ the integral of which in the parameters space is the
`geometrical'
Hannay's angle.

The main purpose of this paper is to study to what extent the above
results, concerning the
canonical equivalence and the identity of the geometric phases of
dissipative and conservative linear
oscillators, can be generalized to nonlinear ones at least in the first
order approximation of the
perturbation theory. In the following we restrict ourselves to the quartic
generalized oscillator and
compare it with the quartic damped oscillator. We do not consider cubic
terms because, at this order, they are non resonant {\it i.e.} without
effect on the phase
equation [8].

The first section is devoted to the quartic generalized oscillator the
Hamiltonian of
which is deduced from (2) by the addition of a term proportional to $Q^4$.
In order to
calculate the geometric phase of this system we extend the method of reduction
to normal forms to the case where the parameters vary slowy with time. We
are then able to solve
perturbatively the Hamilton equations and to find the appropriate canonical
transformations under the
two assumptions of weak nonlinearity and of adiabaticity of the variation
of the parameters. In
particular we determine the adiabatic invariant $I$ of the system and we
find the expression of the
first order nonlinear correction to the geometric Hannay part of the angle
$\Theta$. The reason for
choosing the normal forms technic rather than the averaging method is that,
besides the fact that
it can in principle be developed to any order of perturbation (in the
adiabatic regime), it is a
standard approach for the study of nonlinear dissipative systems; it will
allow us to
deduce the appropriate canonical transformations in the dissipative case
from the above ones.

The second section is devoted to the study of the damped quartic oscillator
the Lagrangian of which
is deduced from (1) by the addition of a term proportional to $q^4$. Using
the time-dependent
canonical transformations found in section 1 and keeping the same
hypotheses of weak nonlinearity and
of adiabaticity we obtain the following results. For finite values of the
damping parameter
$\lambda$, {\it i.e.} when resonance does not exist, the quartic term is
without effect on the
geometric part of the angle. In that case the quartic damped oscillator can
be shown to be
canonically equivalent to the linear generalized harmonic oscillator.
However, there also exists a weak damping limit characterized by a
magnitude of $\lambda$ going to
zero with the adiabatic parameter in such a way that the resonance
phenomenon resurges in the limit.
In this limit the quartic term contributes and one recovers the Hannay's
angle of the quartic generalized oscillator determined in the previous
section. This later result
generalizes to nonlinear systems the result established in [6] for linear
oscillators.

\bigskip

\bigskip

\noindent 1. GENERALIZED QUARTIC OSCILLATOR
\medskip
The simplest nonlinear extension of the generalized harmonic oscillator
leading to a resonant term in the equation of motion is obtained by adding
a quartic term to
the Hamiltonian (2) which thus becomes:  $$H_G (P,Q,\vec \mu) = {P^2 \over
{2}} + \lambda P Q + {{ \omega _o^2} \over 2}  Q^2  + {{\nu } \over 4} Q^4.
\eqno (7)$$
The Hamilton equations for $Q$ and $P$ read
 $$\dot Q = \lambda Q +  P \ \ \ \ , \ \ \   \dot P = -\lambda  P -
\omega_0^2  Q  -\nu Q^3 \  .\eqno (8)$$
In order to solve these nonlinear coupled equations it is convenient to
introduce, in place of $P$
and $Q$, the complex variable $z$, and its complex conjugate $\overline z$,
defined by
$$z =\sqrt{{\omega  \over 2}}\Bigl[ Q - {i\over \omega}\, (\lambda Q +
P )\Bigr ],\ \ \ \ \ \ \  (\omega^2 = \omega_o^2 - \lambda^2) \  . \eqno (9)$$
Equivalently $Q$ and $P$ read: $$ Q={1\over \sqrt {2 \omega}} (z+ \overline
z),\ \ \ \ \  P={{i
\omega - \lambda} \over \sqrt {2 \omega}} z - {{i \omega + \lambda} \over
\sqrt {2 \omega}}\overline z \ . \eqno (10) $$
For time dependent parameters $(\lambda, \omega_0, \nu) \equiv \vec \mu$,
the Hamilton equations for
$Q$ and $P$ lead to the following nonlinear equation for $z$:
 $$\dot z = i\omega  z +{{i \nu} \over {4
\omega ^2}} (z + \overline z)^3 -  { { i \dot \lambda } \over {2 \omega }}
z +  { {\dot \omega
- i \dot \lambda} \over {2 \omega} } \overline z . \eqno (11)$$
(Note that the three parameters $\lambda$, $\omega$ and $\nu$ do not play
the same role: the
time derivative $\dot \nu$ of the parameter associated with the non linear
quartic term does not
appear in (11) in contradistinction with $\dot \lambda$ and $\dot \omega$.)
This equation can be
solved perturbatively using its canonical reduction to  normal form [9] if
one assumes that the
system is weakly nonlinear ($\displaystyle {{\nu \over {\omega^2_o}}\, Q^2
\ll 1}$) and that the
parameters vary adiabatically ($\vec \mu$ is a slowly time varying function
${\vec \mu} (\epsilon t)$
with $\displaystyle {{\epsilon \over \omega_o} \ll 1}$). Under these
conditions, let us introduce
the near to identity transformation
 $$z = u +\delta  \overline u \  \eqno (12) $$
in order to eliminate the nonresonant term  proportional to $\overline u$
into the equation for the
new variable $u$. For $\displaystyle {\delta ={ {\dot \lambda + i \dot
\omega } \over {4 \omega ^2
}}}$ ($\delta$ small, of order $\epsilon$) the coefficient of $\overline u$
cancels and the equation
for $u$ reads:
 $$\dot u = i\Bigl (\omega -  { { \dot \lambda } \over {2 \omega }} \Bigr )
u +{{ i \nu} \over {4
\omega ^2}}( u +  \overline u)^2\Bigl [ (1+ \delta + 3\overline \delta )u +
(1 +4\delta )
 \overline u \Bigr] \  .\eqno (13)  $$
The first term in (13) already exhibits the Hannay's angle of the linear
generalized oscillator.
In order to obtain the non linear correction one must introduce a second
(near to
identity) change of variable  $$u = v +\alpha v^3 + \beta v \overline v ^2
+ \gamma \overline v^3
.\eqno (14)$$  The requirement that the equation for $v$ no longer contains
the nonresonant cubic
terms present in (13) leads to differential equations for the time
dependent coefficients $\alpha$,
$\beta$ and $\gamma$ of the transformation. These differential equations
are explicitly written and
their solutions discussed in the appendix. The resulting equation for $v$,
valid up to the
first order in $\epsilon$ and in the weak nonlinearity parameter's $\nu$, is:
 $$\dot v = i(\omega -{ { \dot \lambda } \over { 2\omega  }}) v +{{3 i
\nu} \over {4 \omega ^2}} (1+{ { \dot \lambda } \over { \omega ^2}}) v
^2\overline v \  \eqno (15) $$
Then, setting $v = A\, e^{i\Theta}$, the equations for the amplitude $A$
and for the angle $\Theta$ read:
 $$  {\dot A }=0,  \eqno (16)$$
 $${\dot \Theta} = \omega \Bigl ( 1 + {{3\nu}\over {4 \omega^3}} A^2 \Bigr
) - {\dot \lambda \over
{2\omega}}\, \Bigl ( 1- {{3\nu} \over {2 \omega^3}}A^2\Bigr ) .\eqno (17) $$
Equation (16) shows that $A$ is an adiabatic invariant related, as  seen
below, to the action $I$ of
the system$$I=   A^2 \  .\eqno(18)$$  The first term in the r.h.s. of
equation (17) accounts for
the well known result that the quartic term $\displaystyle {{\nu \over 4}
Q^4}$ in (7), responsible
of the presence of the resonant $v^2 \overline v$ terms in (15), induces a
renormalization of the
linear frequency $\omega $ which becomes  $$\Omega = \omega (1 + {{3\nu}
\over{4\omega^3}}\,
A^2) \ . \eqno (19)$$   (The facts that the frequency does not explicitly
appear in the expression of
$I$ and that the correction term $\displaystyle {{{3\nu}
\over{4\omega^2}}\, A^2}$ to the linear
frequency is different from the expression found in reference [8] are due
to the use of the amplitude
$A$ of the transformed variable $v$ in (18) and (19) in place of the
amplitude $a$ of the original
variable $Q$; in particular, the usual expression for $\Omega$ in terms of
$a$, $\omega$ and $\nu$
can be recovered from (19) noting that $\displaystyle A = \sqrt {\omega
\over 2}\, a$.) The second
term $\displaystyle {{\dot \lambda \over {2\omega}}\, \Bigl ( -1 + {{3\nu}
\over {2
\omega^2}}A^2\Bigr )}$ in the r.h.s. of equation (17), which exists only
for time dependent
parameters, is the geometric, non integrable, Hannay's part of $\dot
\Theta$. Like the
dynamical part $\Omega$ it also contains a contribution from the nonlinear
resonant term. However, as
this geometrical part is defined up to a total time derivative one can
write it, for $\nu$ constant,
under the same form as in the linear case $\Bigl (\displaystyle {-{\dot
\lambda \over {2\omega}}\Bigr
)}$ in terms of the renormalized frequency $\Omega$ and of a renormalized
damping parameter $\Lambda$
defined by
 $$\Lambda = \lambda \Bigl (1 -{{3\nu} \over {\omega^3}}\, A^2) \ . \eqno (20)$$
Indeed, the equality $\displaystyle{{\dot \Lambda \over {2 \Omega}} = }$
$\displaystyle {{{\dot \lambda \over {2 \omega}}( 1 - {{3\nu} \over
{2\omega^3}}\, A^2)} - {d\over
dt}\Bigl ({{9\nu \lambda} \over {8\omega^4}}\, A^2 \Bigr )}$ is exact up to
the first order in the
weak nonlinear term proportional to $\nu A^2$. Consequently, the equation
(17) for the angle variable
$\Theta$ takes a functional form, $$\dot \Theta = \Omega - {{\dot \Lambda}
\over {2\Omega}} \ , \eqno
(21)$$ identical to the one, $\displaystyle {\dot \Theta = \omega - {{\dot
\lambda} \over
{2\omega}}}$, obtained in the linear case. One can say that, for $\nu$
constant, the effect of the
weak quartic nonlinearity on the phase (the `geometrical' as well as
the`dynamical' parts)
amounts to a simple renormalization of both $\omega$ and $\lambda$.

We now show that the dynamical variables ($I,\Theta$) and
($P,Q$) are related by a time dependent canonical transformation and that
the equations
(16-18) are the Hamilton equations for action-angle variables associated
with the time
dependent Hamiltonian ${\cal H}_G (I,\Theta ,t)$:
 $$ {\cal H}_G (I,\Theta ,t) = I \omega \Bigl ( 1 - {{\dot
\lambda} \over {2\omega ^2}}\Bigr ) + I^2{{3\nu} \over {8\omega^2 }}\Bigl
(1 + {{\dot \lambda} \over
{\omega^2}  } \Bigr ) \,  \ . \eqno (22) $$
This can be verified by inspection of the set of successive transformations
$(P,Q)\to (-i\overline
z, z)\to (iu, \overline u)\to (-i\overline v , v)\to (I,\Theta)$, which
will be also of importance in
section 2. As concerns the first transformation $(P,Q)\to (-i\overline z,
z)$ its generating function
$ F(Q,z)$ is obtained by integration of the equations $\displaystyle{ {
{\partial  F } \over
{\partial  Q}}=P(Q,z)} $ and ${ \displaystyle{{\partial  F } \over
{\partial  z}} =i \overline
z}(Q,z)$ deduced from the differential identity (characterizing a canonical
transformation)
 $$ PdQ -H_G dt=-i {\overline z} dz -{\cal H }_G^z dt +dF\  .\eqno (23)$$
Taking into account the relations (9-10) one gets
$$ F(Q,z) = - {1 \over 2}(i \omega + \lambda ) Q^2 + i \sqrt{2 \omega }
Qz - {1\over 2} i z^2 \  .\eqno (24) $$
The Hamiltonian ${\cal H }_G^z$ for the new conjugate variables
$(-i\overline z,z)$ is then obtained
from the relation $\displaystyle{ {\cal H }_G^z = H_G  + {\partial F \over
\partial t}}$ .
Its expression reads $$ {\cal H }_G^z = \omega  z \overline z +{{ \nu} \over {16
\omega ^2}} (z + \overline z)^4 +  { {i\dot \omega } \over {4 \omega }} ( z
^2 - \overline z ^2) -
 { {\dot \lambda} \over {4 \omega} }(z +  \overline z )^2  \eqno (25)$$
and one can verify that the Hamilton equations
$$ \dot z = { {\partial {\cal H }_G^z } \over {\partial (-i\overline z)}}\
\  ,\ \   -i\dot {\overline
z} = -{ {\partial {\cal H }_G^z  } \over {\partial  z}}\  \eqno (26)$$
indeed coincide with the equation (11) for $z$ and the corresponding one
for $\overline z$.

The transformation $(-i\overline z, z)\to (iu, \overline u)$ associated
with the generating function
$$G(z,\overline u) = -iz{\overline u } + {1 \over 2 } i \delta {\overline u
}^2 - {1 \over 2 } \overline
\delta z^2 \eqno(27) $$
is also canonical. The expression of the Hamiltonian  ${\cal H }_G^u$ for
the new conjugate variables
$(iu,\overline u)$ is obtained from the relation  $\displaystyle{ {\cal H
}_G^u = {\cal H }_G^z
+{\partial G \over \partial t}}$ and, in this hamiltonian formulation of
the reduction to normal
form, $\delta $ and $ \overline \delta $ are determined by the requirement
that ${\cal H }_G^u$ no
longer contains terms proportional to ${\overline u}^2$ and $u^2$. This is
equivalent to the  above
requirement that the equation for $u$ (resp. $\overline u$)  no longer
contains non resonant term
proportional to $ \overline u $ (resp $u$) and it leads to the same value
$\displaystyle {{ {\dot
\lambda + i \dot \omega } \over {4 \omega ^2 }}}$ for $\delta$. Then,
${\cal H }_G^u$  reads $$ {\cal
H }_G^u  = (\omega  -  { {\dot \lambda} \over {2 \omega} } )  u \overline u
+{{ \nu} \over {16 \omega
^2}} (u + \overline u)^4 + {{ \nu} \over {4 \omega ^2}} (u + \overline u)^3
(\delta {\overline u } +
\overline \delta u)\  . \eqno (28)$$(In fact $\displaystyle {{\partial G
\over \partial t}}$ does not
contribute to this expression of ${\cal H}_G^u$ valid up to the first order
in $\epsilon$ since it
only contains terms proportional to $\dot \delta$ and $\dot {\overline
\delta}$ and is thus of order
$\epsilon^2$.)

In the same way, the transformation $(iu, \overline u)\to (-i\overline v,
v)$ associated with the
generating function
$$K(\overline u, v) = iv\overline u +\alpha v^3\overline u + {1 \over 3}
\beta v \overline u ^3 +{1 \over 4}
\gamma \overline u ^4 -{1 \over 4}\overline \gamma v^4 \eqno(29) $$
is canonical. The values of $\alpha$, $\beta$ and $\gamma$, now determined
by imposing that
the Hamiltonian $\displaystyle {{\cal H }_G^v =  {\cal H }_G^u  + {\partial
K\over \partial t}}$ for
the new conjugate variables $(-i\overline v,v)$ does not contain quartic
non-resonant terms, are those
found in the appendix; the expression of ${\cal H }_G^v$  is  $${\cal H }_G^v  =
(\omega  -  { {\dot \lambda} \over {2 \omega} } ) v \overline v+{{ 3\nu}
\over {8 \omega ^2}}(1+ {
{\dot \lambda} \over { \omega^2} }) v^2  \overline v^2\  .   \eqno (30)$$

Finally the transformation $(-i\overline v, v)\to (I,\Theta)$ associated with
the generating function
$$ M (v,\Theta) = -{1 \over 2}v^2 \exp(-2i\Theta) \eqno (31)$$
is canonical and the Hamiltonian (30) transforms into (22) as announced.
This expression (22) shows that $I = A^2$ is the
action of the system and allows to determine the adiabatic invariant as a
function of the energy.
\bigskip
\noindent   2. DAMPED QUARTIC OSCILLATOR

\medskip

The damped quartic oscillator which we consider in this section is
described by the equation:
$$\ddot q  +\omega_0 ^2 q+ 2\lambda \dot q + \nu q^3 =0. \eqno (32) $$  It
can be obtained from the
generalized time-dependent Caldirola-Kanai Lagrangian [10]
 $${\cal L}_D (q,\dot q, \vec \mu) = {1 \over 2} e^{2 \int^t\! \lambda(s)\,
ds }({\dot q}^2
-\omega_0^2 q^2 - {1 \over 2} \nu q^4). \eqno (33)$$
In terms of the new variable
$$ Q = q  e^{\int^t\! \lambda(s)\, ds }\eqno (34)$$
the Lagrangian  (33) reads
 $$ L_D (Q,\dot Q, \vec \mu) = {1 \over 2}({\dot Q}^2 - 2 \lambda Q \dot Q
-\omega^2 Q^2 - {1 \over 2} \nu Q^4   e^{ - 2 \int^t\! \lambda(s)\, ds } )
\eqno (35)$$
and the corresponding time-dependent Hamiltonian $ H_D (P,Q,\vec \mu)$,
where $P = \displaystyle
{{\partial L_D} \over {\partial \dot Q}} = {\dot Q} -\lambda Q$, takes a
form : $$ H_D (P,Q,\vec \mu)
= { P^2 \over 2} + \lambda PQ +{\omega_0^2 \over 2}  Q^2 + {\nu \over 4}
Q^4   e^{ - 2 \int^t\!
\lambda(s)\, ds } \eqno (36) $$ identical to the expression (7) for the
Hamiltonian of the quartic
generalized oscillator, except the crucial exponential factor in the
quartic term. Then, making the
same successive changes of variables as in the previous section, namely
$(P,Q)\to (-i\overline z,
z)\to (iu, \overline u)\to (-i\overline v , v)$, we obtain without any new
calculation the following
Hamiltonian for the conjugate variables $(-i\overline v , v)$:
 $$ {\cal H }_D^v  = (\omega  -  { {\dot \lambda} \over {2 \omega} } ) v
\overline v+{{ 3\nu} \over {8 \omega ^2}}(1+ { {\dot \lambda} \over {
\omega^2} }) v^2  \overline v^2\  e^{ - 2 \int^t\!
\lambda(s)\, ds }\, \  \eqno (37)   $$

At this point a remark is of order. In contradistinction with the
generalized quartic oscillator the
damped quartic oscillator does not exhibit the resonance phenomenon. The
theory of
normal forms teaches that it is possible in that case to eliminate non
linear terms. As a consequence
one expects (as announced in [6]) that the Hannay's angle does no get any
non linear contribution.
Let us see how this comes  in the formalism of canonical transformations.
To this end we introduce
the (near to identity) change of variable  $$v = w + i \sigma w^2 \overline
w   .\eqno (38)$$  The
transformation $(-i\overline v , v)\to (iw, \overline w)$ is canonical for
$\sigma$ real. It
corresponds to the generating function
 $$ N (v,\overline w)  = - i v \overline w - {1 \over 2} \sigma v^2
{\overline w} ^2  \eqno (39) $$
and the transformed Hamiltonian $\displaystyle {{\cal H }_D^w = {\cal H
}_D^v + {{\partial N}\over
{\partial t}}}$ reads
 $$ {\cal H }_D^w  = (\omega  -  { {\dot \lambda} \over {2 \omega} } ) w
\overline w - {1 \over 2}  \Bigl ( \dot \sigma - {{ 3\nu} \over {4 \omega
^2}}(1+
 { {\dot \lambda} \over { \omega^2} })\,  e^{ - 2 \int^t\! \lambda(s)\, ds
} \Bigr ) w^2  \overline
w^2\  . \eqno (40)   $$
The second term in the r.h.s. of (40) can be set equal to zero. Indeed, because
of the presence of the exponential factor the corresponding value of
$\sigma$ remains
small (proportional to $\nu$) which guarantees that the transformation (38)
is close to identity.
Thus for $\lambda$ finite the
quartic damped oscillator is canonically equivalent to the generalized
harmonic oscillator and the phases of the two systems are identical. Note
that such an
elimination can not be done on the Hamiltonian (30) because the integration
of $\dot
\sigma$ would lead to unbounded terms proportional to the time $t$.

Let us now consider the case where the magnitude of the damping parameter
$\lambda$ is close to zero
in such a way that the resonance phenomenon cannot be avoided. More
precisely let
$\lambda $ be of the form  $$\lambda (t) = {\epsilon }^a{\tilde \lambda}
({\epsilon }^b t) \eqno
(41)$$ where $a$ and $b$ are positive numbers such that $a>b>0$ and $a+b
=1$. $b$ positive ensures
the validity of the adiabatic hypothesis for $\lambda$ and $a+b=1$ that
$\dot \lambda$ remains of
order $\epsilon$, like $\dot \omega$. For such a behaviour of $\lambda$ the
integral $\int^t\!
\lambda(s)\, ds $ is of order $\epsilon ^{a-b}$ with $(a-b) >0$. Then,
replacing the exponential
factor in (37) by the relevant terms of its expansion one gets the
following expression (valid up to
the first order in $\epsilon$) for the Hamiltonian ${\cal H }_D^v$: $$
{\cal H }_D^v  = (\omega  -  {
{\dot \lambda} \over {2 \omega} } ) v \overline v+{{ 3\nu} \over {8 \omega
^2}}(1 - 2 \int^t\!
\lambda(s)\, ds +{ {\dot \lambda} \over { \omega^2} }) v^2  \overline v^2 \
.\eqno (42)   $$
For the same reasons as for the Hamiltonian (30) the non linear terms in
(42) cannot be
eliminated. It is interesting to note that the term proportional to the
time derivative of the
parameters, $\displaystyle {-{ {\dot \lambda} \over {2 \omega} } v
\overline v} + { {3\nu \dot \lambda} \over { 8\omega^4} } v^2  \overline
v^2 $, is identical in
(30) and (42). Therefore in this weak damping limit the Hannay's angle
of the quartic damped oscillator is the same as the one of the quartic
generalized harmonic
oscillator. Due to the presence
of the supplementary dynamical term $\displaystyle {-{{ 3\nu} \over {4
\omega ^2}}\int^t\!
\lambda(s)\,ds  \, v^2  \overline v^2}$ in (42) the two systems seem to be
not canonically equivalent.
However one can express the Hamiltonian (42) in terms of the renormalized
parameter $ \tilde \nu =
\nu (1 - 2 \int^t\! \lambda(s)\, ds)$ (a change which, in the degree of
accuracy of our calculations,
does not affect the geometrical nonlinear part of the angle) to get an
expression
$$ {\cal H }_D^v  = (\omega  -  {
{\dot \lambda} \over {2 \omega} } ) v \overline v+{{ 3\tilde \nu} \over {8
\omega ^2}}(1 +{ {\dot
\lambda} \over { \omega^2} }) v^2  \overline v^2 \eqno (43)   $$
similar to (30). The damped quartic oscillator with parameters $\omega_0$,
$\lambda$ and $\nu $
is thus, in this weak damping limit and in the first order approximation of
perturbation theory,
canonically equivalent to the generalized quartic oscillator with
parameters $\omega_0$, $\lambda$
and $\tilde \nu$ and both systems have identical Hannay's angles. These
results are the
generalization for nonlinear oscillators of the result established in [6]
for linear ones.

\bigskip
\noindent APPENDIX
\medskip

The nonlinear, near to identity, change of variable (14) transforms the
equation (13) for $u$
into the following equation for the new variable $v$:
$$\dot v = i\Bigl (\omega -{{\dot \lambda}\over {2\omega}}\Bigr ) v +
{{3i\nu }\over {4\omega ^2}}\bigl
(1 +2 (\delta +\overline \delta )\bigr )\, v^2 \overline v  - D_1 \alpha \
v^3 - D_2 \beta \
v\overline v ^2 - D_3 \gamma \  \overline v ^3 \eqno (A.1)$$
where the differential operators $D_1$, $D_2$ and $D_3$ are such that
 $$D_1 \alpha = \dot \alpha + 2 i\Bigl (\omega -{{\dot \lambda}\over
{2\omega}}\Bigr ) \alpha
-{{i\nu }\over {4\omega ^2}} (1 + \delta  + 3\overline \delta ) \eqno (A.2)$$
$$ D_2 \beta = \dot \beta -  2 i\Bigl (\omega -{{\dot \lambda}\over
{2\omega}}\Bigr )
\beta - {{3i\nu }\over {4\omega ^2}} (1 + 3\delta  + \overline \delta )
\eqno (A.3)$$
$$ D_3 \gamma = \dot \gamma -  4 i\Bigl (\omega -{{\dot \lambda}\over
{2\omega}}\Bigr )
\gamma - {{i\nu }\over {4\omega ^2}} (1 + 4\delta ) \eqno (A.4)$$
For $\alpha$ solution of $D_1 \alpha = 0$, $\beta$ solution of $D_2 \beta =
0$ and $\gamma$ solution of $D_3 \gamma = 0$ the nonresonant cubic terms
can be eliminated in
(A.1). The three differential equations having the same structure, it is
sufficient to study the
behaviour of the solution of one of them, for example $\alpha $. Since
$\displaystyle {\delta =
{{\dot \lambda + i\dot \omega}\over {4\omega ^2}}}$, the differential
equation for $\alpha$ reads $$
\dot \alpha + 2 i\Bigl (\omega -{{\dot \lambda}\over {2\omega}}\Bigr )
\alpha -{{i\nu }\over {4\omega
^2}} (1 + {{2\dot \lambda - i\dot \omega}\over {2\omega ^2}} ) = 0 \
.\eqno (A.5)$$ This equation
which contains terms of order zero and of order one with respect to the
small adiabatic parameter
$\epsilon$ can be solved by a perturbative method and its solution
 can be written under the form of an expansion with respect to $\epsilon$.
Neglecting the
terms of order $\epsilon$ in (A.5) ({\it i.e.} all the time derivatives)
one finds the zero order approximate solution
$\displaystyle {\alpha _0  ={\nu \over {8\omega ^3}}}$. Then putting
$\alpha = \alpha_0 + \alpha_1$ into (A.5) one gets an
equation for $\alpha_1$ which contains terms of order one and of
order two with respect to $\epsilon$. Neglecting the terms proportional to
$\epsilon^2$ in this equation, one obtains an
expression (of order $\epsilon$) for $\alpha_1$ and so on. One verifies
that $\alpha$, and thus also $\beta$ and $\gamma$, are
small for weak nonlinearity and in the adiabatic hypothesis (the leading
terms are proportional to $\nu$ and $\epsilon$). This
validades the above calculations which suppose that the transformation is
near to the identity. Finally the nonresonant cubic
terms can indeed be eliminated in (A.1) which reduces to (15). \vfill
\eject \centerline {REFERENCES}

\bigskip

\noindent [1] HANNAY J-H 1985 J. Phys. A: Math. Gen. {\bf 18} 221

\noindent \ \ \ \  BERRY M V 1985 J. Phys. A: Math. Gen. {\bf 18} 15

\noindent [2] BERRY M V 1984 Proc. R. Soc. A {\bf 392} 45

\noindent [3] KEPLER T B,  KAGAN M L and EPSTEIN I R 1991 Chaos {\bf 1} 455

\noindent [4] LANDSBERG A S 1992 Phys. Rev. Lett. {\bf 69} 865

\noindent [5] TODOROV V Yu and DEBROV V 1993 Phys. Rev. {\bf A 50} 878

\noindent [6] USATENKO O V, PROVOST J P and VALLEE G  1996 J. Phys. A:

\noindent \ \ \ \  Math. Gen. {\bf 29} 2607

\noindent [7] JACKIW R 1988 Int J Mod Phys { \bf A 3 } 285

\noindent [8] LANDAU L and LIFCHITZ E     Physique Th\'eorique Tome 1
M\'ecanique ,

\noindent [9] ARNOLD V I 1978 Mathematical Methods in Classical Mechanics,

\noindent \ \ \ \  Appendix 7 p.385, Springer-Verlag New York.

\noindent \ \ \ \ VERHULST F 1990 Nonlinear Differential Equations and
Dynamical

\noindent \ \ \ \ Systems, Springer-Verlag Berlin Heidelberg.

\noindent [10] CALDIROLA P 1941 Nuovo Cimento {\bf 18} 393

\noindent \ \ \ \ \  KANAI E 1948 Prog. Theor. Phys. {\bf 3} 440

\end{document}